# Random-Order Enumeration for Self-Reducible NP-Problems

Pengyu Chen
pchen.research@gmail.com
Harbin Institute of Technology
Harbin, China

Dongjing Miao
miaodongjing@hit.edu.cn
Harbin Institute of Technology
Harbin, China

Weitian Tong
wtong.research@gmail.com
Georgia Southern University
Statesboro, USA

Zizheng Guo
zguo.research@gmail.com
Harbin Institute of Technology
Harbin, China

Jianzhong Li
lijzh@siat.ac.cn
Shenzhen Institute of Advanced
Technology
Shenzhen, China

Zhipeng Cai
zcai@cs.gsu.edu
Georgia State University
Atlanta, USA

## ABSTRACT

In plenty of data analysis tasks, a basic and time-consuming process is to produce a large number of solutions and feed them into downstream processing. Various enumeration algorithms have been developed for this purpose. An enumeration algorithm produces all solutions of a problem instance without repetition. To be a statistically meaningful representation of the solution space, solutions are required to be enumerated in uniformly random order. This paper studies a set of self-reducible NP-problems in three hierarchies, where the problems are polynomially countable ($\text{Sr}_{\text{NP}}^{\text{FP}}$), admit FPTAS ($\text{Sr}_{\text{NP}}^{\text{FPTAS}}$), and admit FPRAS ($\text{Sr}_{\text{NP}}^{\text{FPRAS}}$), respectively. The trivial algorithm based on a (almost) uniform generator is in fact inefficient. We provide a new insight that the (almost) uniform generator is not the end of the story. More efficient algorithmic frameworks are proposed to enumerate solutions in uniformly random order for problems in these three hierarchies. (1) For problems in $\text{Sr}_{\text{NP}}^{\text{FP}}$, we show a random-order enumeration algorithm with polynomial delay (PDREnum); (2) For problems in $\text{Sr}_{\text{NP}}^{\text{FPTAS}}$, we show a Las Vegas random-order enumeration algorithm with expected polynomial delay (PDLVREnum); (3) For problems in $\text{Sr}_{\text{NP}}^{\text{FPRAS}}$, we devise a fully polynomial delay Atlantic City random-order enumeration algorithm with expected delay polynomial in the input size and the given error probability $\delta$ (FPACREnum), which has a probability of at least $1 - \delta$ becoming a Las Vegas random-order enumeration algorithm. Finally, to further improve the efficiency of the random-order enumeration algorithms, based on the master/slave paradigm, we present a parallelization with 1.5-optimal enumeration delay and running time, along with the theoretical analysis.

## 1 INTRODUCTION

Modern data processing systems have the ability of processing massive data in the areas of data mining [25], social networks [12], bioinformatics [20], cheminformatics [4, 30], etc. In most cases, it needs to produce a large number of solutions and feed them for downstream processing such as heuristic search [5], online aggregation [18, 24], streaming machine learning [28], and query feedback [15, 21]. How to efficiently enumerate all solutions one by one without repetition is widely investigated in the database community, including but not limited to enumerating query answers of a given acyclic conjunctive query [3], listing all $k$-cliques of a graph [11, 31], enumerating minimal triangulations of a graph [8] to help query optimization for databases [29], and enumerating regular document spanners for information extraction [14].

Existing enumeration algorithms barely provide a guarantee for the randomness of the enumeration order. And these algorithms usually produce solutions that are very similar to each other, while returning varied solutions has been identified as an important property in a broad sense [1, 2]. For example, given a free-connex acyclic conjunctive query $Q(\mathbf{x})$, Bagan et al. [3] simply enumerated the assignments for variables $x_1, \ldots, x_n$ of $\mathbf{x}$ in the lexicographic order. In this way, the consecutively enumerated assignments are likely to share an identical prefix, every solution enumerated is usually similar to its adjacent ones, thus leading to the non-uniformity.

To be a statistically meaningful representation of the solution space, the enumeration order needs to be provably random [9], or more specifically, each solution needs to be enumerated uniformly from the set of solutions not yet enumerated. Unfortunately, random-order enumeration has not been studied too much in the literature. For the problems that admit a uniform generator, Capelli et al. [7] discussed a straightforward method that transforms the uniform generator into a sampling-without-replacement algorithm. For each emission, this algorithm generates a candidate solution uniformly by the uniform generator and prevents repetitions by saving all enumerated solutions and discarding the solution if it is already stored. This algorithm is not efficient as lots of generated solutions will be discarded, and the expected delay undergoes hyperbolic growth. To the best of our knowledge, the questions of what kind of problems is tractable in terms of random-order enumeration and how to develop more efficient random-order enumeration algorithms for them are still open.

In this paper, we aim to answer the questions mentioned above. We consider a set of self-reducible enumeration problems ($\text{Sr}_{\text{NP}}$) whose decision version is in NP and can be solved by referring to smaller instances of the same problem. Jerrum et al. [22] developed an almost uniform generator for problems in $\text{Sr}_{\text{NP}}$, whose counting version admits a fully polynomial-time randomized approximation scheme (FPRAS). Even though it can be trivially transformed into a sampling-without-replacement algorithm, it still suffers from the disadvantages mentioned above.

*Contributions.* We provide a new insight into more efficient random-order enumeration, and present efficient algorithmic frameworks avoiding these disadvantages for problems in $\text{Sr}_{\text{NP}}$ whose solutions can be (approximately) counted efficiently. For each emission, a solution is generated exactly from the solutions not yet enumerated instead of from all solutions. Therefore, no solutions will be discarded, and thus the expected delay is lower and does not increase in terms of complexity. We also show our random-order enumeration algorithm frameworks can be parallelized efficiently. Formally, our contributions are listed as follows.

Firstly, if the solutions of a problem in $S_{R_{NP}}$ can be counted exactly in polynomial time, we design an efficient random-order enumeration algorithm that runs in polynomial delay. As the number of solutions for any input can be counted exactly and efficiently, our main idea is to create a mapping between the set of solutions and a set of integers. Then uniformly emitting a solution in one round is equal to picking an integer uniformly from the integer set corresponding to the solutions not yet enumerated.

Secondly, if the solutions of a problem in $S_{R_{NP}}$ can be approximately counted by a fully polynomial-time approximation scheme (Fptas), we design an efficient Las Vegas random-order enumeration algorithm that runs in expected polynomial delay. For each emission, it successfully generates a uniformly random solution from the solutions not yet enumerated with a constant probability and keeps repeating the current generation step if it fails. As the number of solutions for input can only be estimated roughly, our main idea is to create a mapping between the set of solutions and a set of non-overlapping and near-equal-width intervals. To generate a solution uniformly for one emission, we can pick a real number uniformly from the union of the intervals corresponding to the solutions not yet enumerated. We further propose several delicate strategies to realize uniformity efficiently. Some of these strategies may be of independent interest.

Thirdly, if the approximate counting for the solutions of a self-reducible enumeration problem can only succeed by chances, i.e., this problem admits a fully polynomial-time randomized approximation scheme (Fpras), we design an efficient fully polynomial delay Atlantic City random-order enumeration algorithm, which has a high probability becoming a Las Vegas random-order enumeration algorithm. Our main idea is to boost the success rate of the approximate counting and guarantee the consistency of it across multiple runs.

Finally, we parallelize our algorithms on multiple computers based on the master/slave paradigm [26], and provide a theoretical analysis on the bound of the running time and the enumeration delay based on queueing model, which proved that our parallelized algorithm has 1.5-optimal enumeration delay and running time.

Due to space restrictions, proofs and some details of our results are given in the appendix, along with some additional experiments conducted to evaluate the efficiency of our algorithms.

## 2 PRELIMINARIES

In this paper, let $\mathbb{N}$ be the set of all natural numbers, and for any natural numbers $i \leq j$, let $\mathbb{N}[i, j]$ be the set $\{i, i + 1, \ldots, j\}$. We use "$\circ$" to denote the string concatenation operator, and use $\lambda$ to denote the empty string. Given a set of intervals $B$, we abuse the notation $[0, 1) \setminus B$ to denote the set of real numbers in $[0, 1)$ but not in any interval of $B$. For any order $\pi$ over a set $S$ and any integer $i \in \mathbb{N}[1, |S|]$, let $\pi(S)[i]$ be the $i$-th element of $S$ under the order $\pi$.

## 2.1 Enumeration and Polynomial Delay

Fixed an alphabet $\Sigma$, an *enumeration problem* is defined as a pair $(X, Sol)$ where $X$ is the set of inputs, and for any input $x \in X$, $Sol(x)$ is a finite set of solutions for $x$. An *enumeration algorithm* for an enumeration problem is able to produce a sequence of all solutions in $Sol(x)$ without repetition.

**Np-enumeration problem.** An *Np-enumeration problem* $(X, Sol)$ is an enumeration problem if $\forall x \in X$ of size $n$,

(1) $\forall w \in Sol(x)$, the length of $w$ is $poly(n)$,
(2) $\forall w \in \Sigma^*$, it takes $poly(n)$ time to test if $w \in Sol(x)$.

**Polynomial Delay.** As defined in [23], an enumeration algorithm runs in *polynomial delay* if the time between enumerating every two consecutive solutions is at most $poly(n)$.

## 2.2 Uniform Generator

A *uniform generator* for a set $S$ is a randomized algorithm that outputs every element in $S$ with an identical probability. Formally, given a finite set $S$, a randomized algorithm $\mathcal{G}$ is a *uniform generator* for $S$ if $\forall w \in S$, $\Pr(\mathcal{G} \text{ outputs } w) = \frac{1}{|S|}$.

The existence of a uniform generator is not guaranteed for many fundamental problems [2]. We consider a relaxed notation of generator, i.e., *Las Vegas uniform generator*, which either generates an element uniformly or informs about the failure.

*Definition 2.1 (Las Vegas Uniform Generator).* Given a finite set $S$, a randomized algorithm $\mathcal{G}$ is a *Las Vegas uniform generator* if

(1) there exists a constant $c \in (0, 1)$, $\Pr(\mathcal{G} \text{ outputs } false) < c$,
(2) $\forall w \in S$, $\Pr(\mathcal{G} \text{ outputs } w | \mathcal{G} \text{ does not output } false) = \frac{1}{|S|}$.

## 2.3 Random-Order Enumeration

For an enumeration problem $(X, Sol)$, given any input $x \in X$, all solutions in $Sol(x)$ are required to be produced one by one without repetition. Furthermore, if each emission of an enumeration algorithm $\mathcal{M}$ is generated by a uniform generator (Las Vegas uniform generator, respectively) from the remaining solutions, we say $\mathcal{M}$ is a *random-order enumeration algorithm* (Las Vegas random-order enumeration algorithm, respectively), which is abbreviated as REnum (LVREnum, respectively).

*Definition 2.2 (Polynomial Delay Random-Order Enumeration Algorithm).* A randomized algorithm $\mathcal{M}$ is a *polynomial delay random-order enumeration algorithm* (PDREnum) if $\mathcal{M}$ is a REnum with polynomial delay.

*Definition 2.3 (Polynomial Delay Las Vegas Random-Order Enumeration Algorithm).* A randomized algorithm $\mathcal{M}$ is a *polynomial delay Las Vegas random-order enumeration algorithm* (PDLVREnum) if $\mathcal{M}$ is a LVREnum with expected polynomial delay.

*Definition 2.4 (Fully Polynomial Delay Atlantic City Random-Order Enumeration Algorithm).* A randomized algorithm $\mathcal{M}$ is a *fully polynomial delay Atlantic City random-order enumeration algorithm* (FPACREnum) if $\mathcal{M}$ takes $(x, \delta) \in X \times (0, 1)$ as input and produces a sequence of solutions in $Sol(x)$ one by one without repetition such that (1) the probability that $\mathcal{M}$ is LVREnum is at least $1 - \delta$, and (2) the expected time taken by $\mathcal{M}$ to produce every two consecutive solutions is polynomial in $|x|$ and $\log \frac{1}{\delta}$.

## 2.4 Approximate Counting

A *counting algorithm* for an enumeration problem $(X, Sol)$ is used for computing the size of the solution set. Formally, the counting algorithm $\mathcal{A} : X \to \mathbb{N}$ is defined as $\mathcal{A}(x) = |Sol(x)|$ for any $x \in X$. If an enumeration problem admits a polynomial-time counting algorithm $\mathcal{A}$ we say it is *polynomially countable*.

For those enumeration problems that are not polynomially countable, we adopt *approximate counting algorithms* to efficiently approximate the value $|Sol(x)|$. More precisely, a *fully polynomial-time approximation scheme* (Fptas) for an enumeration problem $(X, Sol)$ is an approximate counting algorithm $\mathcal{B} : X \times (0, 1) \to \mathbb{N}$ such that

$$\forall (x, \varepsilon) \in X \times (0, 1), \big| \mathcal{B}(x, \varepsilon) - |Sol(x)| \big| \leq \varepsilon |Sol(x)|$$

and runs in polynomial time of $|x|$ and $\varepsilon^{-1}$.

Similarly, a *fully polynomial-time randomized approximation scheme* (Fpras) for an enumeration problem $(X, Sol)$ is an approximate counting algorithm $C : X \times (0, 1)^2 \to \mathbb{N}$ such that

$$\forall (x, \varepsilon, \delta) \in X \times (0, 1)^2, \Pr\big( |C(x, \varepsilon, \delta) - |Sol(x)|| \leq \varepsilon |Sol(x)| \big) \geq 1 - \delta$$

and runs in polynomial time of $|x|$, $\varepsilon^{-1}$ and $\log \frac{1}{\delta}$.

## 2.5 Self-reducible NP-enumeration Problems

Intuitively, an enumeration problem is *self-reducible* if the set of solutions can be recursively divided into different subsets where each subset contains all the solutions sharing an identical prefix [27], and the enumeration problem can be solved by enumerating these subsets. Formally, a self-reducible enumeration problem can be defined as follows.

*Definition 2.5 (Self-Reducible Enumeration Problem).* An enumeration problem $(X, Sol)$ over an alphabet $\Sigma$ is said to be *self-reducible* if the following conditions hold:

(a) There exists a polynomial-time computable length function $\zeta : X \to \mathbb{N}$ such that $\forall x \in X, \forall w \in Sol(x), |w| = \zeta(x)$.

(b) For any $x \in X$ such that $\zeta(x) = 0$, it takes polynomial time in $|x|$ to check if $Sol(x)$ contains the empty string $\lambda$.

(c) There exists a polynomial-time computable self-reduce function $\Psi : X \times \Sigma^* \to X$ such that $\forall (x, w) \in X \times \Sigma^*$, it satisfies

- $|\Psi(x, w)| \leq |x|$
- $\zeta(\Psi(x, w)) = \max\{\zeta(x) - |w|, 0\}$
- $Sol(\Psi(x, w)) = \{w' | w \circ w' \in Sol(x)\}$

Without loss of generality, we assume $\Sigma = \{0, 1\}$ in the rest of the paper as our main results can be easily extended to the general case. We study the self-reducible NP-enumeration problems ($Sr_{NP}$) and particularly focus on three types of $Sr_{NP}$ shown as below.

(1) $Sr_{NP}^{FP}$, set of problems in $Sr_{NP}$ and polynomially countable,
(2) $Sr_{NP}^{FPTAS}$, set of problems in $Sr_{NP}$ and admits an Fptas,
(3) $Sr_{NP}^{FPRAS}$, set of problems in $Sr_{NP}$ and admits an Fpras.

By definition, the hierarchy follows $Sr_{NP}^{FP} \subseteq Sr_{NP}^{FPTAS} \subseteq Sr_{NP}^{FPRAS}$.

## 3 RANDOM-ORDER ENUMERATION FOR $Sr_{NP}^{FP}$, $Sr_{NP}^{FPTAS}$ AND $Sr_{NP}^{FPRAS}$

In this section, we design the polynomial delay random-order enumeration algorithm (PDREnum), polynomial delay Las Vegas random-order enumeration algorithm (PDLVREnum), and fully polynomial delay Atlantic City random-order enumeration algorithm (FPACREnum) for $Sr_{NP}^{FP}$, $Sr_{NP}^{FPTAS}$, and $Sr_{NP}^{FPRAS}$, respectively.

As every problem in $Sr_{NP}^{FP}$ is polynomially countable, there is a way to count the number of solutions exactly and efficiently. Meanwhile, problems in $Sr_{NP}^{FP}$ are self-reducible. We can order the solutions of $x$ in the lexicographical order and naturally map each solution to an integer in $\mathbb{N}[1, |Sol(x)|]$. To enumerate solutions of $x$ in random order, we simply enumerate an integer in random order. This is our main idea to design a PDREnum for problems in $Sr_{NP}^{FP}$.

For problems in $Sr_{NP}^{FPTAS} \setminus Sr_{NP}^{FP}$, the exact number of solutions for an input $x$ is difficult and we are only able to estimate it as close as possible with an Fptas. The previous idea for problems in $Sr_{NP}^{FP}$ apparently does not work for problems in $Sr_{NP}^{FPTAS}$. Because the problems in $Sr_{NP}^{FPTAS}$ are self-reducible, the solutions of $x$ can be sorted in the lexicographical order even though these solutions are not revealed explicitly. Besides, the self-reduciblity allows us to recursively partition the solution set of any input $x$ into two subsets such that solutions in each subset share the same prefix. We map the set of solutions of $x$ to the interval $[0, 1)$ and each subset of solutions to some sub-interval of $[0, 1)$. While partitioning a solution set, we partition the corresponding interval such that each resultant interval has a length nearly proportional to the estimated number of solutions in the corresponding subset. Eventually, each solution of $x$ will one-to-one correspond to a sub-interval of $[0, 1)$. All these sub-intervals are non-overlapping and have near-equal widths. Therefore, enumerating a solution is equal to enumerating an sub-interval, which can be easily realized by picking a real number uniformly from the sub-intervals corresponding to the solutions not yet enumerated. The near-equal widths of these intervals imply the near-uniform generation for each emission. We then will apply a "redo" strategy to force each emission uniform with respect to the solutions not yet enumerated. This is our main idea to design the PDLVREnum for problems in $Sr_{NP}^{FPTAS}$.

Consider the problem in $Sr_{NP}^{FPRAS} \setminus Sr_{NP}^{FPTAS}$. The number of solutions for any input $x$ can only be estimated accurately with a probability of at least $1 - \delta$, where $\delta \in (0, 1)$. We can apply the previous ideas for $Sr_{NP}^{FPTAS}$. Additionally, we need to boost the probability of providing high estimate accuracy of the number of solutions, which can be done by running the algorithm multiple times. This is our main idea to design the FPACREnum for problems in $Sr_{NP}^{FPRAS}$.

## 3.1 Efficient PDREnum for $Sr_{NP}^{FP}$

Before describing the PDREnum algorithm for problems in $Sr_{NP}^{FP}$, we here introduce an important operation *random-access* [9], which is formally defined as follows.

*Definition 3.1 (Random-Access).* Given an enumeration problem $(X, Sol)$ (with length function $\zeta$ and self-reduce function $\Psi$) and its counting algorithm $\mathcal{A}$ that outputs value $|Sol(x)|$ for any $x \in X$. Let $\pi$ be a fixed order, a *random-access* (RAccess$^\pi$ for short) is an algorithm developed based on $\mathcal{A}$ and $\pi$ such that for any $(x, i) \in X \times \mathbb{N}^+$, RAccess$^\pi(x, i, \zeta(x), \Psi, \mathcal{A})$ returns $\pi(Sol(x))[i]$ if and only if $i \in \mathbb{N}[0, |Sol(x)|]$. Otherwise, it returns *false*.

**PDREnum.** As shown in [9], if a random-access is implemented efficiently, a PDREnum algorithm follows immediately for problems in $Sr_{NP}^{FP}$. Concretely, if an enumeration problem $(X, Sol)$ admits a random-access RAccess$^\pi$ in polynomial time, a PDREnum for problem $(X, Sol)$ can be implemented as follows.

---

**Algorithm 1:** ARA

---

**Input:** $x, \zeta, \Psi, \mathcal{A}$

1   $N_x \leftarrow \mathcal{A}(x)$;   //the value of $|Sol(x)|$
2   $d \leftarrow \zeta(x)$;   //the length of solutions in $Sol(x)$
3   **for** $k = 1$ **to** $N_x$ **do**
4      randomly enumerate an integer $i \in \mathbb{N}[1, N_x]$;
5      $w \leftarrow$ RAccess$^\pi(x, i, d, \Psi, \mathcal{A})$;
6      **output** $w$;

---

In each loop of steps 3-5, Fisher-Yates Shuffle [13] is employed to enumerate integers in uniformly random order in constant delay. As shown in line 4, whenever an integer $i$ is enumerated, ARA (algorithm 1) employs RAccess$^\pi(x, i, \zeta(x), \Psi, \mathcal{A})$ to compute $\pi(Sol(x))[i]$ and outputs it. ARA terminates after $|Sol(x)|$ loops, its performance depends on the running time of the random-access.

**Efficient implementation of RAccess$^\pi$.** It is able to develop a polynomial time random-access for every problem in $\text{Sr}_{\text{NP}}^{\text{FP}}$, which implies the first result of this paper that $\text{Sr}_{\text{NP}}^{\text{FP}}$ admits PDREnum. Intuitively, computing $\pi(Sol(x))[i]$ is polynomially tractable when $\pi$ is the lexicographical order. This is because if an enumeration problem $(X, Sol)$ is self-reducible and polynomially countable, then for any input $x \in X$, it takes polynomial time to count the number of solutions sharing the same prefix, so that for any position $k$, the $k$-th bit $w[k]$ of $\pi(Sol(x))[i]$ can be figured out by its prefix $w[1]w[2]\ldots w[k-1]$ in a very efficient way.

---

**Algorithm 2:** RAccess$^{\pi^*}$

---

**Input:** $x, i, d, \Psi, \mathcal{A}$
**Output:** $\pi^*(Sol(x))[i]$

1   **if** $i > \mathcal{A}(x)$ **then return** $false$;
2   **if** $d = 0$ **then return** empty string $\lambda$;
3   $N_x(0) \leftarrow \mathcal{A}(\Psi(x, 0))$;
4   **if** $i \leq N_x(0)$ **then return** $0 \circ$ RAccess$^{\pi^*}(\Psi(x, 0), i, d-1, \Psi, \mathcal{A})$;
5   **else return** $1 \circ$ RAccess$^{\pi^*}(\Psi(x, 1), i - N_x(0), d-1, \Psi, \mathcal{A})$;

---

Formally, let $\pi^*$ be the lexicographical order. If a problem $(X, Sol)$ in $\text{Sr}_{\text{NP}}$ admits a polynomial time counting algorithm $\mathcal{A}$, then a random-access for it can be implemented as algorithm 2. The random-access runs recursively similar with the binary search: RAccess$^{\pi^*}(x, i, \zeta(x), \Psi, \mathcal{A})$ first compute the number of the solutions starting with bit 0, say $N_x(0)$, by employing the counting algorithm $\mathcal{A}$. In order to figure out if $\pi(Sol(x))[i]$ starts with 0, it tests whether $i$ exceeds $N_x(0)$. If $i$ does not exceed $N_x(0)$, it can be decided that $\pi(Sol(x))[i]$ starts with 1. Otherwise, it starts with 0. Once a bit is figured out, it recursively decides the following bits.

**Time complexity of RAccess$^{\pi^*}$.** Obviously, the recursion depth of RAccess$^{\pi^*}$ is at most $\zeta(x)$, hence $\mathcal{A}$ runs at most $\zeta(x)$ times during the process of RAccess$^{\pi^*}$. For any input $x$ of size $n$ of a problem in $\text{Sr}_{\text{NP}}$, there exists a length function $\zeta$ and a self-reduce function $\Psi$. Let $T_\Psi(n)$ be the time of computing $\Psi(x)$, $T_\zeta(n)$ be the time of calculating $\zeta(x)$, and $T_{\mathcal{A}}(n)$ be the running time of $\mathcal{A}(x)$, then RAccess$^{\pi^*}$ runs in time $O(\zeta(x) \cdot (T_\Psi(n) + T_{\mathcal{A}}(n)))$.

**Delay of ARA (algorithm 1).** It has been shown that Fisher-Yates Shuffle runs in constant delay, that is, the delay of ARA depends on only the execution time of the random-access, thus yielding the following result.

**Theorem 3.2.** *If an enumeration problem $(X, Sol)$ is in $\text{Sr}_{\text{NP}}^{\text{FP}}$, given its polynomial-time computable length function $\zeta$, its polynomial-time computable self-reduce function $\Psi$ and its polynomial-time counting algorithm $\mathcal{A}$, then ARA (algorithm 1) is a PDREnum with $O(\zeta(x) \cdot (T_\Psi(n) + T_{\mathcal{A}}(n)))$ delay after $O(T_\zeta(n) + T_{\mathcal{A}}(n))$ time preprocessing for any input $x \in X$ of size $n$.*

## 3.2 Efficient PDLVREnum for $\text{Sr}_{\text{NP}}^{\text{FPTAS}}$

In contrast to $\text{Sr}_{\text{NP}}^{\text{FP}}$, it is impossible to develop a polynomial-time random-access for any problem in $\text{Sr}_{\text{NP}}^{\text{FPTAS}} \setminus \text{Sr}_{\text{NP}}^{\text{FP}}$. This is because it is hard to test whether the input integer $i$ exceeds $|Sol(x)|$ when it cannot be exactly counted efficiently [9]. However, every such problem admits an approximate counting algorithm of arbitrary accuracy. In the following, we show an efficient way to take a good use of the Fptas for problems in $\text{Sr}_{\text{NP}}^{\text{FPTAS}}$.

### 3.2.1 An Efficient PDLVREnum.
When there is a way to count the number of solutions exactly and efficiently, the PDREnum ARA maps each solution of input $x$ to an integer in $\mathbb{N}[1, |Sol(x)|]$. To enumerate solutions in random order, one can simply enumerate integers in random order. However, for problems in $\text{Sr}_{\text{NP}}^{\text{FPTAS}} \setminus \text{Sr}_{\text{NP}}^{\text{FP}}$, there is a way to count the solutions only approximately, the previous idea does not work.

**Enumerating intervals instead of integers.** Instead of enumerating integers from $\mathbb{N}[1, |Sol(x)|]$, we enumerate intervals from $[0, 1)$. Specifically, we partition $[0, 1)$ into $|Sol(x)|$ intervals near-equal in width, and map each solution $w$ to one of the intervals. Different solutions are not allowed to be mapped to an identical interval. Then, whenever a number $r$ is picked uniformly from $[0, 1)$, if $r$ is in some interval $I$ and the solution $w$ mapped to $I$ has not yet been enumerated, then enumerate $w$. The probability of the picked number belonging to interval $I$ is proportional to the length $|I|$.

Formally, we are able to develop a PDLVREnum for any enumeration problem $(X, Sol)$ where for any $x \in X$ of size $n$ there exist a surjection $f: [0, 1) \rightarrow Sol(x)$, named *shift function*, and a *seed generator* $\mathcal{G}$ such that

$f$:   $\forall w \in Sol(x)$, $\{r \in [0, 1) | f(r) = w\}$ is an interval of length between $\frac{2}{3} \frac{1}{|Sol(x)|}$ and $\frac{4}{3} \frac{1}{|Sol(x)|}$, meanwhile, $\forall r \in [0, 1)$, $f(r)$ and $f^{-1}(f(r))$ can be computed in $poly(n)$ time,

$\mathcal{G}$:   for any disjoint interval set $B$, outputs a random number $r$ as a seed uniformly from $[0, 1) \setminus B$ in $poly(n)$ time.

The key idea is to compute $f$ and $f^{-1}$ efficiently without listing the solution set $Sol(x)$ completely in advance, and to compute $\mathcal{G}$ efficiently without traversing all the intervals in $B$.

**Interval-access for $f$ and $f^{-1}$.** Every time given a seed $r \in [0, 1)$, both the solution $f(r)$ and its interval $f^{-1}(f(r))$ need to be figured out. As shown in Section 3.2.2, we develop a polynomial time algorithm *interval-access* compute both $f(r)$ and $f^{-1}(f(r))$ for any given seed $r \in [0, 1)$.

*Definition 3.3 (Interval-Access).* For any enumeration problem $(X, Sol)$, given a real number $r \in [0, 1]$, interval-access (IACCESS for short) returns the solution $f(r)$ along with its interval $f^{-1}(f(r))$.

**Seed generator $\mathcal{G}$.** Duplicate solutions are not allowed to be enumerated, thus needing to prevent the generation of real numbers from an identical interval. Accordingly, once a solution $w$ is enumerated, it is necessary to ban the interval $f^{-1}(w)$, only real numbers from the intervals not yet banned, say available intervals, are allowed to be picked next time. To this end, as shown in Section 3.2.3, we show an efficient implementation of $\mathcal{G}$ based on AVL tree designed to well organize the set of banned intervals.

**PDLVRENUM.** We here claim that there exist efficient implementations of interval-access and generator $\mathcal{G}$ which are detailed in sections 3.2.2-3.2.3. Then, for any problem $(X, Sol)$ in $\text{SR}_{NP}^{FPTAS}$ (with length function $\zeta$ and self-reduce function $\Psi$), given its FPTAS $\mathcal{B}$ which is able to provide an approximation of the value $|Sol(x)|$ with a relative error $\varepsilon$ for any input $(x, \varepsilon) \in X \times (0, 1)$, a polynomial delay Las Vegas random-order enumeration algorithm (PDLVRENUM) can be formally shown as algorithm 3.

---

**Algorithm 3:** AIA

**Input:** $x, \zeta, \Psi, \mathcal{B}$

1  $B \leftarrow \emptyset$; //the set of intervals already banned
2  $L \leftarrow 1$; //the sum of lengths of available intervals
3  $d \leftarrow \zeta(x)$; //the length of solution
4  $\tilde{N}_x(\lambda) \leftarrow \mathcal{B}(x, \frac{1}{3})$; //the estimated number of solutions
5  $\varphi^* \leftarrow \frac{4}{9} \cdot \frac{1}{\tilde{N}_x(\lambda)}$; //the non-uniformity correction factor
6  **while** $L > 0$ **do**   //exist a solution not yet enumerated
7     $r \leftarrow \mathcal{G}(B, L)$;
8     $\{f(r), f^{-1}(f(r))\} \leftarrow \text{IACCESS}(x, r, \lambda, 0, 1, d, \Psi, \mathcal{B})$;
9     pick a number $p \in [0, 1)$ randomly and uniformly;
10   **if** $p \geq \frac{\varphi^*}{|f^{-1}(f(r))|}$ **then continue**; //redo the generation
11   **output** $f(r)$;   //enumerate $w$ if it passes the test
12   update $B \leftarrow B \cup \{f^{-1}(f(r))\}$;   //ban $f^{-1}(f(r))$
13   $L \leftarrow L - |f^{-1}(f(r))|$;   //shrink the available space

---

In AIA (algorithm 3), every time $\mathcal{G}$ picks a seed $r \in [0, 1)$ from an interval not yet banned, so that IACCESS produces a solution without repetition in each loop. With the help of steps 8 and 9, solutions are finally enumerated uniformly. Until $L = 0$, all solutions in $Sol(x)$ are enumerated without omission.

**Delay of AIA (algorithm 3) and its uniformity.** AIA consists of preprocessing phase (*i.e.*, steps 1-5) and enumerating phase (steps 6-13). The delay of AIA depends mainly on the performance of the enumerating phase. We here show that AIA is actually a PDLVRENUM for $\text{SR}_{NP}^{FPTAS}$, which is the second result of this paper.

**THEOREM 3.4.** *Given enumeration problem $(X, Sol)$ in $\text{SR}_{NP}^{FPTAS}$, let $\zeta$, $\Psi$ and $\mathcal{B}$ be the polynomial computable length function, self-reduce function and FPTAS of $(X, Sol)$. For any $x \in X$ of size $n$ and any $\varepsilon \in (0, 1)$, let $T_\zeta(n)$, $T_\Psi(n)$ and $T_\mathcal{B}(n, \varepsilon^{-1})$ be the running times of $\zeta(x)$, $\Psi(x)$ and $\mathcal{B}(x, \varepsilon)$. Given an $O(\zeta(x) \cdot (T_\Psi(n) + T_\mathcal{B}(n, \zeta(x))))$-time interval-access and an $O(\zeta(x))$-time seed generator $\mathcal{G}$, along with a well-organized banned interval set $B$ which*

*requires $O(\zeta(x))$-time to insert a new interval, AIA (algorithm 3) is a PDLVRENUM with expected $O(\zeta(x) \cdot (T_\Psi(n) + T_\mathcal{B}(n, \zeta(x))))$ delay after $O(T_\zeta^*(n) + T_\mathcal{B}(n, 1))$ time preprocessing.*

In the rest of this section, an efficient implementation of interval-access IACCESS is formally described in Section 3.2.2. and a seed generator $\mathcal{G}$ based on AVL tree is presented in Section 3.2.3.

### 3.2.2 *Efficient implementation of Interval-Access (IACCESS).*

Given an input $x \in X$ and a seed $r \in [0, 1)$, interval-access is designed to compute the solution $w = f(r)$ and its interval $f^{-1}(w)$. We concretely define a proper $f$ whose inverse $f^{-1}$ maps each solution $w \in Sol(x)$ to an interval of a width nearly $\frac{1}{|Sol(x)|}$. In the following, we first formally describe the shift function $f$ and its inverse $f^{-1}$, then provide an algorithm to compute them efficiently without listing all solutions of $Sol(x)$ in advance.

**The shift function $f$ and its inverse $f^{-1}$.** Given any partition of $[0, 1)$ consisting of $|Sol(x)|$ intervals $I_1, \dots, I_{|Sol(x)|}$, we can list all solutions of $Sol(x)$ in lexicographical order and define $f(I_k)$ as the $k$-th the solution $w_k$ ($\pi^*(Sol(x))[k]$), where

$$\forall 1 \leq k \leq |Sol(x)|, \quad f^{-1}(w_k) = \left[ \sum_{i=1}^{k-1} |I_i|, \sum_{i=1}^{k} |I_i| \right).$$

In fact, in this way, even without listing all solutions in advance, $f$ and $f^{-1}$ can be still computed in a recursive way similar with binary search. We next detail the recursive computation of $f$ and its inverse $f^{-1}$ after showing a near-uniform partition of $[0, 1)$.

*Near-uniform partition of $[0, 1)$.* The method here is *recursive partition* which employs an FPTAS to recursively split the current interval into two parts, until the depth is up to $\zeta(x)$, $[0, 1)$ is finally partitioned into $|Sol(x)|$ intervals nearly equal in width. Ideally, a pivot $p$ is first calculated and split $[0, 1)$ into $P_0 = [0, p)$ and $P_1 = [p, 1)$ such that $\forall b \in \{0, 1\}$, $P_b = \bigcup_{\{w \in \Sigma^* | b \circ w \in Sol(x)\}} f^{-1}(b \circ w)$, which implies $|P_b| = \frac{|\{w \in \Sigma^* | b \circ w \in Sol(x)\}|}{|Sol(x)|}$. Recursively, pivots of $P_0$ and $P_1$ are calculated, then split them again. Until the recursion depth is up to $\zeta(x)$, we get $|Sol(x)|$ equal-width intervals finally.

However, without listing all the solutions in $Sol(x)$, it is hard to calculate the pivot precisely for every current interval. Instead, we compute a quasi-pivot by estimating the ideal size of every interval. Formally, given an FPTAS $\mathcal{B}(x, \varepsilon)$ for enumeration problem $(X, Sol)$ in $\text{SR}_{NP}$, $\forall x \in X$, $\forall w \in Sol(x)$, let $d = \zeta(x)$, let $w[1, i]$ as the $i$-length prefix of $w$, and $\tilde{N}_x(w[1, i])$ be the estimation of $|Sol(\Psi(x, w[1, i]))|$ computed by $\mathcal{B}$ in the condition of $\varepsilon = \frac{1}{8d}$. During the recursion, whenever splitting an interval $I = [l, h)$ corresponding to a prefix $w[1, k]$ of $w$, the quasi-pivot is defined as

$$\tilde{p} = l + \frac{\tilde{N}_x(w[1, k] \circ 0)}{\tilde{N}_x(w[1, k] \circ 0) + \tilde{N}_x(w[1, k] \circ 1)}(h - l).$$

**LEMMA 3.5.** $\forall x \in X$, $\forall w \in Sol(x)$, *when the recursive partition based on quasi-pivot terminates, it follows*

(1) $|f^{-1}(w)| = \prod_{i=0}^{d-1} \frac{\tilde{N}_x(w[1, i+1])}{\tilde{N}_x(w[1, i] \circ 0) + \tilde{N}_x(w[1, i] \circ 1)}$,

(2) $\frac{2}{3} \frac{1}{|Sol(x)|} \leq |f^{-1}(w)| \leq \frac{4}{3} \frac{1}{|Sol(x)|}$.

**IACCESS computes $f$ and $f^{-1}$ recursively.** Based on lemma 3.5, we present algorithm 4 shown as below. Similar with binary search,

IAccess recursively decides the value of each bit of the solution $w$ such that $r \in f^{-1}(w)$. Every time it calculates the current quasi-pivot $\tilde{p}$, figures out which side seed $r$ belongs to and shrinks the interval, finally obtains the solution and its interval.

---

**Algorithm 4: IAccess**

**Input:** $x, r, w', l, h, d, \Psi, \mathcal{B}$
**Output:** $\{w, f^{-1}(w)\}$ such that $r \in f^{-1}(w)$
1 **if** $d = 0$ **then return** $\{w', [l, h)\}$;
2 $w_0 \leftarrow w' \circ 0, \quad w_1 \leftarrow w' \circ 1$;
3 $\tilde{N}_x(w_0) \leftarrow \mathcal{B}(\Psi(x, w_0), \frac{1}{8d}), \quad \tilde{N}_x(w_1) \leftarrow \mathcal{B}(\Psi(x, w_1), \frac{1}{8d})$;
4 $\tilde{p} \leftarrow l + \frac{\tilde{N}_x(w_0)}{\tilde{N}_x(w_0) + \tilde{N}_x(w_1)}(h - l)$;
5 **if** $r < \tilde{p}$ **then return** IAccess$(x, r, w_0, l, \tilde{p}, d - 1, \Psi, \mathcal{B})$;
6 **else return** IAccess$(x, r, w_1, \tilde{p}, h, d - 1, \Psi, \mathcal{B})$;

---

**Performance of IAccess shown as algorithm 4.** For any input $x \in X$ of size $n$, IAccess shown as algorithm 4 calls $\mathcal{B}$ at most $2n$ times. $\mathcal{B}$ runs in $T_{\mathcal{B}}(n, \zeta(x))$ which is polynomial in $n$, thus yielding the following result.

LEMMA 3.6. *For any $x \in X$ of size $n$, IAccess shown as algorithm 4 runs in $O(\zeta(x)(T_\Psi(n) + T_{\mathcal{B}}(n, \zeta(x))))$ time.*

#### 3.2.3 Efficient Implementation of Seed Generator $\mathcal{G}$.

The task of generator $\mathcal{G}$ is to pick a real number from $[0, 1) \setminus B$ randomly and uniformly. In general, $[0, 1) \setminus B$ is not an interval. This results in a failure of the trivial generator working on an interval. Therefore, we next provide a seed generator $\mathcal{G}$ working on a union of disjoint intervals.

Given $B = \{I_i = [l_i, h_i) | i \in \mathbb{N}[1, |B|]\}$ which is the set of disjoint intervals already banned, w.l.o.g., assume $h_i \le l_{i+1}$ for every $i \in \mathbb{N}[1, |B| - 1]$. Let $L = 1 - \sum_{i=1}^{|B|} |I_i|$. Our seed generator $\mathcal{G}$ works as below,

(1) pick a real number $y \in [0, L)$ randomly and uniformly
(2) compute offset $b = \sum_{i=1}^{k^*} |I_i|$ such that $h_{k^*} \le y + b < l_{k^*+1}$
(3) return $y + b$

In a naïve way computing the offset in step (2), it needs to sort and traverse $B$ in order to find $k^*$, thus turning to an exponential-time algorithm. In order to develop a polynomial-time generator $\mathcal{G}$, it is necessary to well organize set $B$ so as to speed up interval search and insertion.

**Optimization based on AVL tree.** We employ AVL tree to enable efficient search and update so as to speed up $\mathcal{G}$ along with the maintenance of $B$ in step 12 of algorithm 3. Given $B$, we build an AVL tree $T$ consisting of $|B|$ nodes for $B$ such that

(1) Each node $u$ stores a unique interval $I = [l, h) \in B$ by maintaining $u.l$ and $u.h$;
(2) Each node $u$ maintains $u.left$ and $u.right$ pointing to its left child and right child, and if $v$ is the left (right) child of $u$, then $v.h \le u.l$ ($u.h \le v.l$);
(3) Each node $u$ maintains $u.take$ which is the sum of lengths of the intervals in the subtree rooted at $u$, that is,

$$u.take = \begin{cases} u.h - u.l, & \text{if } u \text{ is leaf,} \\ u.h - u.l + u.left.take + u.right.take, & \text{otherwise.} \end{cases}$$

(4) The height of $T$ is $O(\log |B|)$.

Obviously, after each interval insertion, the time to maintain an AVL tree is $O(\log |B|)$ time which is at most $O(\zeta(x))$.

*The computation of offset $b$.* In order to compute the exact offset for $y$, a binary search on the AVL tree $T$ is carried out by accumulating offset in a top-down fashion. At each node $u$, if $I_u$ is higher than the target $I_{k^*}$, then down to the left child to look for increment of the offset, instead, if $I_u$ is lower than the target $I_{k^*}$, then update the offset by at least $|I_u|$ and down to the right child to test if the offset should increase again. The time for computing $b$ is linear to the height of $T$ which is $O(\zeta(x))$.

**Seed Generator $\mathcal{G}$.** Based on the optimization mention above, a fast seed generator $\mathcal{G}$ is provided as algorithm 5.

---

**Algorithm 5: $\mathcal{G}$**

**Input:** $T, L$
**Output:** seed $r \in [0, 1) \setminus B$
1 pick a real number $y \in [0, L)$ randomly and uniformly;
2 $b \leftarrow 0, temp \leftarrow 0, u \leftarrow root(T)$;
3 **while** $u \ne nil$ **do**
4 $\quad$ **if** $u.left = nil$ **then** $temp \leftarrow 0$; **else** $temp \leftarrow u.left.take$;
5 $\quad$ **if** $(y + b) + temp < u.l$ **then** $u \leftarrow u.left$;
6 $\quad$ **else** $b \leftarrow b + temp + (u.h - u.l), u \leftarrow u.right$;
7 **return** $y + b$

---

LEMMA 3.7. *For any disjoint interval set $B$, the seed generator $\mathcal{G}$ shown as Algorithm 5 runs in $O(\zeta(x))$ time.*

### 3.3 Efficient FPACREnum for $\text{Sr}_{\text{NP}}^{\text{FPRAS}}$

In this section, we consider enumeration problems in $\text{Sr}_{\text{NP}}$ that admit an FPRAS. FPRAS is a randomized algorithm which may return inconsistent answers across multiple runs on an identical input. We here show an FPACREnum obtained by replacing the FPTAS with an FPRAS in AIA and IAccess (algorithm 3 and 4) while providing the guarantee of consistent estimations across multiple runs along with a tight bound of relative error.

**Keep consistency by dictionary.** Given an input $x \in X$, build a dictionary $D : \Sigma^* \to \mathbb{N}$ for every solution $w$ to save $\tilde{N}_x(w')$ for any possible prefix $w'$ of $w$. Specifically, if an estimation $\tilde{N}_x(w')$ returned by the FPRAS is already in $D$, say $D[w']$, then we reuse it without estimating it again. The size of the dictionary is $O(|B|)$ in total which is linear to the number of the enumerated solutions, so that a binary search yields an $O(\log |B|) = O(\zeta(x))$ running time.

**FPACREnum.** Formally, the fully polynomial delay Atlantic City random-order enumeration algorithm AXA with input $(x, \delta, \zeta, \Psi, C)$ is shown as algorithm 6, where $\delta \in (0, 1)$ is the error probability, *i.e.* the probability that AXA is LVREnum is at least $1 - \delta$.

**Interval-access with dictionary.** As shown in Algorithm 7, XAccess is an interval-access that guarantees consistent partitions across multiple runs with a dictionary $D$. For any $(x, r) \in X \times [0, 1)$, the shift function $f(r)$ and its inverse $f^{-1}(f(r))$ are computed by XAccess such that $\frac{2}{3} \frac{1}{|Sol(x)|} \le |f^{-1}(f(r))| \le \frac{4}{3} \frac{1}{|Sol(x)|}$ with high probability. The dictionary $D$ is constantly updated by XAccess.

---

**Algorithm 6: AXA**

---

**Input:** $x, \delta, \zeta, \Psi, C$

1   $B \leftarrow \emptyset, D \leftarrow \emptyset, L \leftarrow 1, d \leftarrow \zeta(x), \tilde{N}_x(\lambda) \leftarrow C(x, \frac{1}{3})$;

2   $\varphi^* \leftarrow \frac{4}{9} \frac{1}{\tilde{N}_x(\lambda)}$;

3   **while** $L > 0$ **do**

4      $r \leftarrow \mathcal{G}(B, L)$;

5      $\{f(r), f^{-1}(f(r)), D\} \leftarrow \text{XAccess}(x, r, \delta, \lambda, 0, 1, d, \Psi, C, D)$;

6      pick a number $p \in [0, 1)$ randomly and uniformly;

7      **if** $\frac{\varphi^*}{|f^{-1}(f(r))|} \in (\frac{1}{4}, 1)$ **and** $p \geq \frac{\varphi^*}{|f^{-1}(f(r))|}$ **then continue**;

8      **output** $f(r)$; $B \leftarrow B \cup \{f^{-1}(f(r))\}$;

9      $L \leftarrow L - |f^{-1}(f(r))|$;

---

**Algorithm 7: XAccess**

---

**Input:** $x, r, \delta, w', l, h, d, \Psi, C, D$
**Output:** $\{w, f^{-1}(w), D\}$ such that $r \in f^{-1}(w)$

1   **if** $d = 0$ **then return** $\{w', [l, h], D\}$;

2   $w_0 \leftarrow w' \circ 0$,    $w_1 \leftarrow w' \circ 1$;

3   **if** $D[w_0] = nil$ **then** $D[w_0] \leftarrow C(\Psi(x, w_0), \frac{1}{8d}, 2^{-(d+1)}\delta)$;

4   **if** $D[w_1] = nil$ **then** $D[w_1] \leftarrow C(\Psi(x, w_1), \frac{1}{8d}, 2^{-(d+1)}\delta)$;

5   $\tilde{p} \leftarrow l + \frac{D[w_0]}{D[w_0] + D[w_1]}(h - l)$;

6   **if** $r < \tilde{p}$ **then return** $\text{XAccess}(x, r, w_0, l, \tilde{p}, d - 1, \Psi, C, D)$;

7   **else return** $\text{XAccess}(x, r, w_1, \tilde{p}, h, d - 1, \Psi, C, D)$;

---

**Bound of error probability of AXA (algorithm 3).** AXA is obtained by replacing FPRAS with FPTAS in AIA and IAccess while maintaining a dictionary. In this way, AXA is in fact a FPACRENUM.

**Theorem 3.8.** *Given any enumeration problem* $(X, Sol)$ *in* $\text{SR}_{\text{NP}}^{\text{FPRAS}}$*, let* $\zeta, \Psi$ *and* $C$ *be the polynomial computable length function, self-reduce function and* FPRAS*. For any* $x \in X$ *of size* $n$*, and* $\forall \varepsilon, \delta \in (0, 1)$*, let* $T_\zeta(n), T_\Psi(n)$ *and* $T_C(n, \varepsilon^{-1}, \log \frac{1}{\delta})$ *be the running times of* $\zeta(x)$*,* $\Psi(x)$ *and* $C(x, \varepsilon, \delta)$*. Problem* $(X, Sol)$ *admits fully polynomial delay Atlantic City random-order enumeration algorithm with an expected* $O\left(\zeta(x) \cdot \left(T_\Psi(n) + T_C\left(n, \zeta(x), \zeta(x) \log \frac{1}{\delta}\right)\right)\right)$ *delay.*

**Remarks\*.** For problems in $\text{SR}_{\text{NP}}^{\text{FPRAS}}$, previous results shown in [7] imply an alternative random-order enumeration algorithm, which employs the almost uniform generator given in [22] to generate candidate solutions, and eliminates duplications by discarding those the same with previously enumerated (or output). The time complexity of the uniform generator is the same with the enumeration delay of our algorithm. However, as discussed in [7], after $i$ solutions are enumerated, it requires $O\left(\frac{|Sol(x)|}{|Sol(x)| - i}\right)$ times candidate generation on average, until the $(i + 1)$-th solution is enumerated. Worst still, it is hard to know if all solutions in $Sol(x)$ have been enumerated, even if all runs of FPRAS succeed. In fact, given $\delta \in (0, 1)$, it requires excessively generating $O\left(\left(\zeta(x) + \log \frac{1}{\delta}\right)|Sol(x)|\right)$ candidate solutions in total to ensure the probability that all solutions are enumerated is at least $1 - \delta$.

# 4   PARALLEL ALGORITHM FOR $\text{SR}_{\text{NP}}^{\text{FPRAS}}$

To further improve the efficiency of algorithms proposed in section 3, a parallelization is developed in this section together with a theoretical analysis on the delay based on queueing model. Intuitively, ARA and AIA are both special cases of AXA, hence, we here only show the parallelization of AXA for the sake of simplicity.

## 4.1   Cluster and Master/Slave Paradigm

This section is built on cluster computing systems. A cluster consists of several processor nodes connected by a communication network delivering messages among the nodes with the same capacity. Every node is treated as a black box made of its private memories, processors and communication devices. The interaction of processor nodes is based on the master/slave (M/S) paradigm, where one node acts as the master, all other nodes act as the slaves. The master serves as the control and communication hub.

To solve the problems studied in this paper, the master is responsible for assigning an enumeration workload to each slave and collecting the solutions enumerated by slaves. Each slave runs AXA to enumerate solutions and sends them to master directly.

## 4.2   Parallelization of AXA

A straightforward way is that slaves run AXA directly in parallel and sent enumerated solutions to master which deduplicates and outputs them. However, in this way, there may be a large number of duplicate emissions leading to high communication cost.

### 4.2.1   *Partitioning, dictionary consistency and uniformity.*

To avoid unnecessary network traffic, workload assigned to each slave is required to be disjoint with others. To this end, for $m$ slaves, master partitions $[0, 1)$ into $m$ disjoint ranges $[l_1 = 0, h_1), ..., [l_m, h_m = 1)$ near-equal in width and broadcasts them. Each slave $i$ enumerates solutions from the set $\{w \in Sol(x)|f^{-1}(w) \subset [l_i, h_i)\}$ only, thus preventing duplications. Meanwhile, this static mapping is a near-uniform partitioning sufficient to guarantee load balance. In this partitioning strategy, since $m \ll |Sol(x)|$ in general, master is able to generate $m$ ranges by grouping the $|Sol(x)|$ intervals specified by the shift function $f$, where $f$ is implicitly defined by XAccess. However, two problems remain to be solved.

*Dictionary consistency.* Caused by the partitioning, every two slaves assigned with adjacent ranges may share common prefixes, thus requiring the same approximate counts. To deal with this situation, master computes all the approximate counts that may be shared by slaves with adjacent ranges in advance and sends to each slave its own piece. During the process of enumeration, every slave only looks up its private dictionary to prevent inconsistency without synchronization between itself and the others.

*Guarantee uniformity of enumeration.* A straightforward FIFO master is not able to enumerate solutions uniformly. Instead, master maintains a queue $Q_i$ for each slave $i$. Whenever a solution $w$ from slave $i$ is received, master enqueues $w$ into $Q_i$. Master picks a random number $i \in \mathbb{N}[1, m]$ with a probability proportional to the available range left and dequeues $w$ from $Q_i$ as an output.

### 4.2.2   *A two-phase algorithm.*

As shown in algorithm 8 and 9, preparations to prevent duplicate emissions and inconsistent dictionaries are carried out in Phase-I, then all the solutions are uniformly enumerated in Phase-II.

---

**Algorithm 8: ParaAXA: Master**

**Input:** $m, x, \delta, \zeta, \Psi, C$
// Phase-I: Initialization
1  $d \leftarrow \zeta(x); D \leftarrow \emptyset; l_1 \leftarrow 0;$
2  **for** $i \in \mathbb{N}[1, m-1]$ **do**
3  $\quad \{w, [l, h), D\} \leftarrow \text{XAccess}(x, \frac{i}{m}, \delta, \lambda, 0, 1, d, \Psi, C, D);$
4  $\quad P_i \leftarrow [l_i, h); l_{i+1} \leftarrow l;$
5  $\quad$ **for** $j \in \mathbb{N}[0, \zeta(x) - 1]$ **do** // w.l.o.g., let $w[1,0] = \lambda$
6  $\quad\quad w_0 \leftarrow w[1, j] \circ 0, w_1 \leftarrow w[1, j] \circ 1;$
7  $\quad\quad D_i[w_0] \leftarrow D[w_0], D_{i+1}[w_0] \leftarrow D[w_0];$
8  $\quad\quad D_i[w_1] \leftarrow D[w_1], D_{i+1}[w_1] \leftarrow D[w_1];$
9  $P_m \leftarrow [l_m, 1);$
10  **foreach** $i \in \mathbb{N}[1, m]$ **do** $L_i \leftarrow |P_i|;$
11  $\tilde{N}_x(\lambda) \leftarrow C(x, \frac{1}{3}, 2^{-(d+1)\delta}), \varphi^* \leftarrow \frac{4}{9} \cdot \frac{1}{\tilde{N}_x(\lambda)};$
12  **foreach** $i \in \mathbb{N}[1, m]$ **do send** $\{x, \delta, \varphi^*, P_i, D_i\}$ to slave $i;$
// Phase-II: Enumeration
13  **while** $\sum_{j=1}^{m} L_j > 0$ **do**
14  $\quad$ randomly pick $i \in \mathbb{N}[1, m]$ with probability $p_i = \frac{L_i}{\sum_{j=1}^{m} L_j};$
15  $\quad$ dequeue $\{w, f^{-1}(w)\}$ from the queue $Q_i$ and **output** $w;$
16  $\quad L_i \leftarrow L_i - |f^{-1}(w)|;$

---

**Phase-I: Initialization.** For each slave $i$, master generates a message consisting of input $x$, error probability $\delta$, interval $[l_i, h_i)$, uniformity correction factor $\varphi^*$ and dictionary $D_i$, and sends it to $i$.

*Interval partitions.* Master employs XAccess to generate an interval as the workload assigned to each slave. For each slave $i \in \mathbb{N}[1, m-1]$, its workload is the solution set $M_i = \{w|f^{-1}(w) \subset [l_i, h_i)\}$ where $l_i$ is the lower endpoint of $f^{-1}(f(\frac{i-1}{m}))$ and $h_i$ is the lower endpoint of $f^{-1}(f(\frac{i}{m}))$, and the workload assigned to slave $m$ is the solutions set $M_m = \{w|f^{-1}(w) \subset [l_m, 1)\}$ where $l_m$ is the lower endpoint of $f^{-1}(f(\frac{m-1}{m}))$. As a result, workloads are pair-wise disjoint, thus preventing duplication, meanwhile, the union of them is exactly $Sol(x)$. W.h.p., $\forall i \in \mathbb{N}[1, m], |h_i - \frac{i}{m}| < |f^{-1}(f(\frac{i}{m}))| < \frac{4}{3} \frac{1}{|Sol(x)|}$ so that w.h.p. $||[l_i, h_i)| - \frac{1}{m}| < \frac{8}{3} \frac{1}{|Sol(x)|}$.

*Generate dictionaries.* Master computes approximate counts possibly shared by slaves with adjacent ranges, then generates dictionary $D_i$ for each slave $i$ and sends to $i$ its own piece. Specifically, $\forall i \in \mathbb{N}[1, m-1]$, let $w$ be solution $f(\frac{i}{m})$, only slaves $i$ and $i+1$ may share prefixes of $w$, i.e., only approximate counts in $\{D[w[1, j] \circ 0]|j \in \mathbb{N}[0, \zeta(x) - 1]\} \cup \{D[w[1, j] \circ 1]|j \in \mathbb{N}[0, \zeta(x) - 1]\}$ are required to be put into $D_i$ and $D_{i+1}$. The size of each $D_i$ is $O(\zeta(x))$.

**Phase-II: Enumeration.** Each slave $i$ enumerates solutions in $\{w|f^{-1}(w) \subset [l_i, h_i)\}$ by running AXA independently. Once a solution $w$ is enumerated by slave $i$, the set $\{w, f^{-1}(w)\}$ is sent to master. After that, $f^{-1}(w)$ is banned. Master constantly picks a slave randomly with the probability proportional to the sum of lengths of intervals corresponding to the solutions not yet output by master, i.e., $\Pr(i \text{ is picked}) = \frac{L_i}{\sum_{j=1}^{m} L_j}$. Once slave $i$ is picked, master dequeues a solution $w$ from $Q_i$ as an output, and reduce $L_i$ by $|f^{-1}(w)|$ for slave $i$. Then, the following theorem follows.

**Theorem 4.1.** *The two-phase algorithm shown as algorithms 8 and 9 is a Las Vegas random-order enumeration algorithm for enumeration problems in $\text{Sr}_{\text{NP}}^{\text{FPRAS}}$ with high probability.*

---

**Algorithm 9: ParaAXA: $i$-th Slave**

**Input:** $\zeta, \Psi, C$
// Phase-I: Initialization
1  **receive** $x, \delta, \varphi^*, l_i, h_i, D_i;$
2  $B_i \leftarrow \{[0, l_i), [h_i, 1)\}; L \leftarrow h_i - l_i; d \leftarrow \zeta(x);$
// Phase-II: Enumeration
3  **while** $L > 0$ **do**
4  $\quad r \leftarrow \mathcal{G}(B_i, L);$
5  $\quad \{f(r), f^{-1}(f(r)), D_i\} \leftarrow$
$\quad \text{XAccess}(x, r, \delta, \lambda, 0, 1, d, \Psi, C, D_i);$
6  $\quad$ pick a number $p \in [0, 1)$ randomly and uniformly;
7  $\quad$ **if** $p \geq \frac{\varphi^*}{|f^{-1}(f(r))|}$ **then continue**;
8  $\quad$ **send** $\{f(r), f^{-1}(f(r))\}$ to the queue $Q_i$ of Master;
9  $\quad B \leftarrow B \cup \{f^{-1}(f(r))\}; L \leftarrow L - |f^{-1}(f(r))|;$

---

### 4.3 Running Time and Enumeration Delay

The running time and delay of our parallel enumeration algorithm are affected by several factors, e.g., the enumeration delay of each slave and the transmission time of solutions. W.l.o.g., assume that

- A1: Each slave enumerates solutions with a fixed delay $s$.
- A2: Solutions from an identical slave are received by master with a fixed delay $t$ where $\frac{t}{s} \leq 1 + o(1)$.
- A3: The time master spent on picking a queue and dequeueing a solution is at least $m$ times less than $t$.
- A4: $m \ll |Sol(x)|$.

**Running time.** Consider each slave $i$, A4 implies w.h.p. $|[l_i, h_i)|$ is very close to $\frac{1}{m}$, thus yielding a w.h.p. upper bound of $|M_i|$ which is $|M_i| < \frac{1}{m}(\frac{2}{3} \frac{1}{|Sol(x)|})^{-1} = \frac{3}{2} \frac{|Sol(x)|}{m}$, since w.h.p. $f^{-1}(w) > \frac{2}{3} \frac{1}{|Sol(x)|}$ for each $w \in Sol(x)$. Therefore, the running time of the parallel algorithm is reduced by at least $\frac{2}{3}m$ times.

**Theorem 4.2.** *Let $T$ be the running time of AXA in a machine. Under A1-A4, the two-phase algorithm takes at most $\frac{3}{2} \frac{T}{m}(1 + o(1))$ time to enumerate all the solutions with high probability.*

**Enumeration Delay.** To prevent the worst case, for any $\alpha \in (0, 1)$, at the beginning of phase-I, fix a parameter $Q$, the master first collects solutions from slaves, and it starts enumeration until each queue has at least $Q$ solutions. It follows that solutions are finally enumerated with $(1 + \alpha)\frac{3}{2} \frac{t}{m}$ delay with high probability.

**Theorem 4.3.** *Given input $(m, x, \delta), \forall \alpha, \delta^* \in (0, 1)$, when $Q$ is $\Theta(\frac{m}{\alpha^2} \log \frac{m}{\alpha \delta^*})$, the time between any two solutions consecutively output by master is $(1+\alpha)\frac{3}{2} \frac{t}{m} = (1+\alpha)\frac{3}{2} \frac{s}{m}(1+o(1))$ with probability at least $1 - \delta - \delta^*$ under assumptions A1-A4.*

*Remarks[*]*. In the best case, the delay of the two-phase algorithm is at most $(1+\alpha)\frac{t}{m}$ with high probability. That is, the performance of the two-phase algorithm is between $(1+\alpha)\frac{t}{m}$ and $(1+\alpha)\frac{3}{2} \frac{t}{m}$. Moreover, the assumption A1 is overly strict. Benefiting from dictionaries, the enumeration delay of each slave decreases with time gradually by getting rid of the recomputation of approximate counts.

## 5 CONCLUSION

We developed random-order enumeration algorithm frameworks for $\text{Sr}_{\text{NP}}^{\text{FP}}$, $\text{Sr}_{\text{NP}}^{\text{FPTAS}}$ and $\text{Sr}_{\text{NP}}^{\text{FPRAS}}$ respectively. The results give a new insight into random-order enumeration that we can do better than the sampling-without-replacement method, which is quite straightforward. Algorithmic frameworks proposed in this paper are much more efficient and the expected delays do not increase in terms of complexity. The parallelization still provides a 1.5-optimal bound of the running time and enumeration delay with high probability.

An interesting topic of future research is to enumerate solutions in random order for other problems whose solutions can be uniformly generated efficiently. And the parallelization may be further improved for different distributed systems.

## REFERENCES

[1] Serge Abiteboul, Gerome Miklau, Julia Stoyanovich, and Gerhard Weikum. 2016. Data, Responsibly (Dagstuhl Seminar 16291). *Dagstuhl Reports* 6, 7 (2016), 42–71. https://doi.org/10.4230/DagRep.6.7.42

[2] Marcelo Arenas, Luis Alberto Croquevielle, Rajesh Jayaram, and Cristian Riveros. 2021. #NFA Admits an FPRAS: Efficient Enumeration, Counting, and Uniform Generation for Logspace Classes. *J. ACM* 68, 6, Article 48 (oct 2021), 40 pages. https://doi.org/10.1145/3477045

[3] Guillaume Bagan, Arnaud Durand, and Etienne Grandjean. 2007. On Acyclic Conjunctive Queries and Constant Delay Enumeration. In *Computer Science Logic*, Jacques Duparc and Thomas A. Henzinger (Eds.). Springer Berlin Heidelberg, Berlin, Heidelberg, 208–222.

[4] Dominique Barth, Olivier David, Franck Quessette, Vincent Reinhard, Yann Strozecki, and Sandrine Vial. 2015. Efficient Generation of Stable Planar Cages for Chemistry. In *Experimental Algorithms*, Evripidis Bampis (Ed.). Springer International Publishing, Cham, 235–246.

[5] Anne Berry, Jean R. S. Blair, Pinar Heggernes, and Barry W. Peyton. 2004. Maximum Cardinality Search for Computing Minimal Triangulations. *Algorithmica* 39, 4 (may 2004), 287–298. https://doi.org/10.1007/s00453-004-1084-3

[6] John Burkardt. 2014. Data for the 01 Knapsack Problem. (August 2014). https://people.sc.fsu.edu/~jburkardt/datasets/knapsack_01/knapsack_01.html

[7] Florent Capelli and Yann Strozecki. 2019. Incremental delay enumeration: Space and time. *Discrete Applied Mathematics* 268 (2019), 179–190. https://doi.org/10.1016/j.dam.2018.06.038

[8] Nofar Carmeli, Batya Kenig, and Benny Kimelfeld. 2017. Efficiently Enumerating Minimal Triangulations. In *Proceedings of the 36th ACM SIGMOD-SIGACT-SIGAI Symposium on Principles of Database Systems* (Chicago, Illinois, USA) *(PODS '17)*. Association for Computing Machinery, New York, NY, USA, 273–287. https://doi.org/10.1145/3034786.3056109

[9] Nofar Carmeli, Shai Zeevi, Christoph Berkholz, Benny Kimelfeld, and Nicole Schweikardt. 2020. Answering (Unions of) Conjunctive Queries Using Random Access and Random-Order Enumeration. In *Proceedings of the 39th ACM SIGMOD-SIGACT-SIGAI Symposium on Principles of Database Systems* (Portland, OR, USA) *(PODS'20)*. Association for Computing Machinery, New York, NY, USA, 393–409. https://doi.org/10.1145/3375395.3387662

[10] Herman Chernoff. 1952. A Measure of Asymptotic Efficiency for Tests of a Hypothesis Based on the sum of Observations. *The Annals of Mathematical Statistics* 23, 4 (1952), 493–507. http://www.jstor.org/stable/2236576

[11] Maximilien Danisch, Oana Balalau, and Mauro Sozio. 2018. Listing K-Cliques in Sparse Real-World Graphs". In *Proceedings of the 2018 World Wide Web Conference* (Lyon, France) *(WWW '18)*. International World Wide Web Conferences Steering Committee, Republic and Canton of Geneva, CHE, 589–598. https://doi.org/10.1145/3178876.3186125

[12] Yon Dourisboure, Filippo Geraci, and Marco Pellegrini. 2009. Extraction and Classification of Dense Implicit Communities in the Web Graph. *ACM Trans. Web* 3, 2, Article 7 (apr 2009), 36 pages. https://doi.org/10.1145/1513876.1513879

[13] Richard Durstenfeld. 1964. Algorithm 235: Random Permutation. *Commun. ACM* 7, 7 (jul 1964), 420. https://doi.org/10.1145/364520.364540

[14] Fernando Florenzano, Cristian Riveros, Martin Ugarte, Stijn Vansummeren, and Domagoj Vrgoc. 2018. Constant Delay Algorithms for Regular Document Spanners. In *Proceedings of the 37th ACM SIGMOD-SIGACT-SIGAI Symposium on Principles of Database Systems* (Houston, TX, USA) *(PODS '18)*. Association for Computing Machinery, New York, NY, USA, 165–177. https://doi.org/10.1145/3196959.3196987

[15] Konstantin Golenberg, Benny Kimelfeld, and Yehoshua Sagiv. 2008. Keyword Proximity Search in Complex Data Graphs. In *Proceedings of the 2008 ACM SIGMOD International Conference on Management of Data* (Vancouver, Canada) *(SIGMOD '08)*. Association for Computing Machinery, New York, NY, USA, 927–940.

[16] Parikshit Gopalan, Adam Klivans, Raghu Meka, Daniel Štefankovic, Santosh Vempala, and Eric Vigoda. 2011. An FPTAS for #Knapsack and Related Counting Problems. In *2011 IEEE 52nd Annual Symposium on Foundations of Computer Science*. 817–826. https://doi.org/10.1109/FOCS.2011.32

[17] Evan Greene and Jon A Wellner. 2017. Exponential bounds for the hypergeometric distribution. *Bernoulli: official journal of the Bernoulli Society for Mathematical Statistics and Probability* 23, 3 (2017), 1911.

[18] Peter J. Haas and Joseph M. Hellerstein. 1999. Ripple Joins for Online Aggregation. In *Proceedings of the 1999 ACM SIGMOD International Conference on Management of Data* (Philadelphia, Pennsylvania, USA) *(SIGMOD '99)*. Association for Computing Machinery, New York, NY, USA, 287–298. https://doi.org/10.1145/304182.304208

[19] Wassily Hoeffding. 1994. *Probability Inequalities for sums of Bounded Random Variables*. Springer New York, New York, NY, 409–426. https://doi.org/10.1007/978-1-4612-0865-5_26

[20] Fereydoun Hormozdiari, Petra Berenbrink, Nataša Pržulj, and Cenk Sahinalp. 2007. Not All Scale Free Networks Are Born Equal: The Role of the Seed Graph in PPI Network Emulation. In *Systems Biology and Computational Proteomics*, Trey Ideker and Vineet Bafna (Eds.). Springer Berlin Heidelberg, Berlin, Heidelberg, 1–13.

[21] Vagelis Hristidis and Yannis Papakonstantinou. 2002. Chapter 58 - Discover: Keyword Search in Relational Databases. In *VLDB '02: Proceedings of the 28th International Conference on Very Large Databases*, Philip A. Bernstein, Yannis E. Ioannidis, Raghu Ramakrishnan, and Dimitris Papadias (Eds.). Morgan Kaufmann, San Francisco, 670–681. https://doi.org/10.1016/B978-155860869-6/50065-2

[22] Mark R. Jerrum, Leslie G. Valiant, and Vijay V. Vazirani. 1986. Random generation of combinatorial structures from a uniform distribution. *Theoretical Computer Science* 43 (1986), 169–188. https://doi.org/10.1016/0304-3975(86)90174-X

[23] David S. Johnson, Mihalis Yannakakis, and Christos H. Papadimitriou. 1988. On generating all maximal independent sets. *Inform. Process. Lett.* 27, 3 (1988), 119–123. https://doi.org/10.1016/0020-0190(88)90065-8

[24] Feifei Li, Bin Wu, Ke Yi, and Zhuoyue Zhao. 2019. Wander Join and XDB: Online Aggregation via Random Walks. *ACM Trans. Database Syst.* 44, 1, Article 2 (jan 2019), 41 pages. https://doi.org/10.1145/3284551

[25] Sowmya R and Suneetha K R. 2017. Data Mining with Big Data. In *2017 11th International Conference on Intelligent Systems and Control (ISCO)*. 246–250. https://doi.org/10.1109/ISCO.2017.7855990

[26] Sartaj Sahni and George Vairaktarakis. 1996. The master-slave paradigm in parallel computer and industrial settings. *Journal of Global Optimization* 9, 3 (1996), 357–377.

[27] Claus-Peter Schnorr. 1976. OPTIMAL ALGORITHMS FOR SELF-REDUABLE PROBLEMS. (1976).

[28] W. Nick Street and YongSeog Kim. 2001. A Streaming Ensemble Algorithm (SEA) for Large-Scale Classification. In *Proceedings of the Seventh ACM SIGKDD International Conference on Knowledge Discovery and Data Mining* (San Francisco, California) *(KDD '01)*. Association for Computing Machinery, New York, NY, USA, 377–382. https://doi.org/10.1145/502512.502568

[29] Susan Tu and Christopher Ré. 2015. DunceCap: Query Plans Using Generalized Hypertree Decompositions. In *Proceedings of the 2015 ACM SIGMOD International Conference on Management of Data* (Melbourne, Victoria, Australia) *(SIGMOD '15)*. Association for Computing Machinery, New York, NY, USA, 2077–2078. https://doi.org/10.1145/2723372.2764946

[30] J. R. Ullmann. 1976. An Algorithm for Subgraph Isomorphism. *J. ACM* 23, 1 (jan 1976), 31–42. https://doi.org/10.1145/321921.321925

[31] Kaijie Zhu, George Fletcher, Nikolay Yakovets, Odysseas Papapetrou, and Yuqing Wu. 2019. Scalable Temporal Clique Enumeration. In *Proceedings of the 16th International Symposium on Spatial and Temporal Databases* (Vienna, Austria) *(SSTD '19)*. Association for Computing Machinery, New York, NY, USA, 120–129. https://doi.org/10.1145/3340964.3340987

# A   DETAILED PROOFS

## A.1   Proof of Theorem 3.4

Proof. According to the definitions of the interval-access and number-generation, with the help of step 6, algorithm 3 enumerates all solutions in $Sol(x)$ without omission obviously. The preprocessing phase takes $O(T_\zeta(n) + T_\mathcal{B}(n, 1))$ time. It remains to prove the uniformity of enumeration and the enumeration delay.

*Uniformity of enumeration.* Let $S$ be the set of solutions not yet enumerated. Consider each loop of steps 7-13, $\mathcal{G}$ pick a number $r$ uniformly from $[0, 1) \setminus B$ in step 7, hence any candidate solution $f(r)$ where $f^{-1}(f(r)) \notin B$ is returned by IAccess in step 7 with the probability

$$\frac{|f^{-1}(f(r))|}{\sum_{w' \in S} |f^{-1}(f(w'))|}.$$

Every candidate solution $f(r)$ returned by IAccess in step 8 passes the test in step 10 with the probability

$$\frac{\varphi^*}{|f^{-1}(f(r))|}.$$

It is in fact a probability. By the definitions of Fptas and $f$, we have $\mathcal{B}(x, \frac{1}{3}) \geq \frac{2}{3}|Sol(x)|$ and $|f^{-1}(f(r))| > \frac{2}{3} \frac{1}{|Sol(x)|}$, thus yielding

$$\frac{\varphi^*}{|f^{-1}(f(r))|} = \frac{1}{|f^{-1}(f(r))|} \frac{1}{\frac{9}{4}\mathcal{B}(x, \frac{1}{3})} < \frac{1}{\frac{2}{3}|Sol(x)|} \frac{1}{\frac{9}{4}\frac{2}{3}\frac{1}{|Sol(x)|}} = 1.$$

In total, each solution $w \in S$ is output in step 11 with an identical probability

$$\frac{|f^{-1}(w)|}{\sum_{w' \in S} |f^{-1}(f(w'))|} \cdot \frac{\varphi^*}{|f^{-1}(w)|} = \frac{\varphi^*}{\sum_{w' \in S} |f^{-1}(f(w'))|}.$$

Therefore, every solution enumerated by algorithm 3 is uniformly generated from the set of solutions not yet enumerated.

*Delay of algorithm 3.* According to the condition, the time cost of step 12 is $O(\zeta(x))$. Thus, the time cost of each loop of steps 7-13 is mainly decided by $\mathcal{G}$ and IAccess. Because IAccess runs in $O(\zeta(x) \cdot (T_\Psi(n) + T_\mathcal{B}(n, \zeta(x))))$ time and $\mathcal{G}$ runs in $O(\zeta(x))$ time, the time cost of each loop of steps 7-13 is $O(\zeta(x) \cdot (T_\Psi(n) + T_\mathcal{B}(n, \zeta(x))))$.

In each loop of steps 7-13, any candidate $f(r)$ returned by IAccess is finally output in step 11 with probability

$$\frac{\varphi^*}{|f^{-1}(f(r))|} > \frac{1}{\frac{4}{3}|Sol(x)|} \frac{1}{\frac{9}{4}\frac{4}{3}\frac{1}{|Sol(x)|}} = \frac{1}{4}.$$

Hence, within 4 loops of steps 7-13, a solution is expected to be enumerated, thus yielding an expected $O(\zeta(x) \cdot (T_\Psi(n) + T_\mathcal{B}(n, \zeta(x))))$ delay. □

## A.2   Proof of Lemma 3.5

Proof. Equation (1) follows immediately since the quasi-pivot based recursive partition is performed along the prefix of $w$ independently. For inequation (2), according to the definition of Fptas, it follows

$$|\tilde{N}_x(w) - |Sol(\Psi(x, w))|| \leq \frac{1}{8d}|Sol(\Psi(x, w))|,$$

let $M = |Sol(x)|$, thus yielding a upper bound of $\tilde{\varphi}$ as below

$$|f^{-1}(w)| = \prod_{i=0}^{d-1} \frac{\tilde{N}_x(w[1, i+1])}{\tilde{N}_x(w[1, i] \circ 0) + \tilde{N}_x(w[1, i] \circ 1)}$$

$$\leq \prod_{i=0}^{d-1} \left(\frac{1 + \frac{1}{8d}}{1 - \frac{1}{8d}}\right) \frac{|Sol(\Psi(x, w[1, i+1]))|}{|Sol(\Psi(x, w[1, i] \circ 0))| + |Sol(\Psi(x, w[1, i] \circ 1))|}$$

$$= \prod_{i=0}^{d-1} \left(\frac{1 + \frac{1}{8d}}{1 - \frac{1}{8d}}\right) \frac{|Sol(\Psi(x, w[1, i+1]))|}{|Sol(\Psi(x, w[1, i]))|} \leq \left(\frac{1 + \frac{1}{8d}}{1 - \frac{1}{8d}}\right)^d \frac{1}{M}$$

$$\leq \left(1 - \frac{1}{8d}\right)^{-2d} \frac{1}{M} \leq \frac{64}{49} \frac{1}{M} < \frac{4}{3} \frac{1}{|Sol(x)|}$$

since $\forall x \in (0, 1), 1 + x < \frac{1}{1-x}$. Similarly, it implies the lower bound of $|f^{-1}(w)|$ as below

$$|f^{-1}(w)|$$

$$\geq \prod_{i=0}^{d-1} \left(\frac{1 - \frac{1}{8d}}{1 + \frac{1}{8d}}\right) \frac{|Sol(\Psi(x, w[1, i+1]))|}{|Sol(\Psi(x, w[1, i] \circ 0))| + |Sol(\Psi(x, w[1, i] \circ 1))|}$$

$$\geq \left(\frac{1 - \frac{1}{8d}}{1 + \frac{1}{8d}}\right)^d \frac{1}{M} \geq \left(1 - \frac{1}{8d}\right)^{2d} \frac{1}{M} \geq \frac{49}{64} \frac{1}{M} > \frac{2}{3} \frac{1}{|Sol(x)|}$$

□

## A.3   Proof of Theorem 3.8

Proof. Consider Fpras $C : X \times (0, 1)^2 \to \mathbb{N}$ of problem $(X, Sol)$, $\forall x \in X, \forall \varepsilon, \delta \in (0, 1), C(x, \varepsilon, \delta)$ returns an estimation of $|Sol(x)|$, say $\tilde{N}_x(\lambda)$, such that $\Pr(|\tilde{N}_x(\lambda) - |Sol(x)|| \geq \varepsilon|Sol(x)|) \geq 1 - \delta$, and runs in a time polynomial of $n$, $\varepsilon^{-1}$ and $\log \frac{1}{\delta}$.

According to the definition of self-reducible enumeration problems, for any solution and every possible prefix of it, $C$ runs at most $\sum_{i=0}^{d} 2^i = 2^{d+1} - 1$ times in XAccess, and runs exactly one time in AXA while computing $N_x$. Let the failure probability of $C$ be $2^{-(d+1)}\delta$, then the probability that all the $2^{d+1}$ runs of $C$ consecutive succeed is at least

$$(1 - 2^{-(d+1)}\delta)^{2^{d+1}} \geq 1 - \delta.$$

If all the $2^{d+1}$ runs of $C$ succeed, then AXA is a LVRenum, and for any $r \in [0, 1)$, $f^{-1}(f(r))$ computed by XAccess satisfies $\frac{\varphi^*}{|f^{-1}(f(r))|} \in (\frac{1}{4}, 1)$. As in step 7, if $\frac{\varphi^*}{|f^{-1}(f(r))|} \notin (\frac{1}{4}, 1)$, which means there must be some run of $C$ failed, we enumerate the current solution $f(r)$ anyway. The purpose is to bound the expected number of loops (at steps 4-9) within 4 before a new solution is enumerated. Since $C$ runs at most $O(\zeta(x))$ times for each candidate solution returned by XAccess, and $\log \frac{2^{d+1}}{\delta} = O(\zeta(x) \log \frac{1}{\delta})$, AXA is FPACRenum with probability greater than $1 - \delta$ and the enumeration delay is $O\left(\zeta(x) \left(T_\Psi(n) + T_C\left(n, \zeta(x), \zeta(x) \log \frac{1}{\delta}\right)\right)\right)$. □

## A.4   Proof of Theorem 4.1

Proof. By the proof of Theorem 3.8, all the runs of Fpras succeed with high probability. It remains to prove that under this condition, every solution output by master is uniformly generated from the set of solutions not yet output, say $S$. It consists of solutions in the $m$ queues and those not yet enumerated by slaves. Let $S_i$ be the set of solutions in $Q_i$ and those not yet enumerated by slave $i$ but in its workload. Obviously, $L_i = \sum_{w \in S_i} |f^{-1}(w)|$. As shown previously, with high probability, $\forall i \in \mathbb{N}[1, m]$, any solution in $S_i$ is enumerated by slave $i$ with probability

$$\frac{\varphi^*}{\sum_{w \in S_i} |f^{-1}(w)|},$$

where $\varphi^*$ is the uniformity correction factor same as in AXA. Since

$$\Pr(i \text{ is picked}) = \frac{L_i}{\sum_{j=1}^{m} L_j} = \frac{\sum_{w \in S_i} |f^{-1}(w)|}{\sum_{w \in S} |f^{-1}(w)|},$$

any solution in $S$ is output by master at step 15 with an identical probability

$$\frac{\varphi^*}{\sum_{w \in S_i} |f^{-1}(w)|} \frac{\sum_{w \in S_i} |f^{-1}(w)|}{\sum_{w \in S} |f^{-1}(w)|} = \frac{\varphi^*}{\sum_{w \in S} |f^{-1}(w)|}.$$

□

### A.5 Proof of Theorem 4.3

PROOF. Whenever a solution $w \in M_i$ is dequeued from $Q_i$, the value of $L_i$ is reduced by $|f^{-1}(w)|$. At any given moment, $L_i = |[l_i, h_i)| - \sum_{w \in M_i \setminus S_i} |f^{-1}(w)|$. Let $E$ be the event that $\forall w \in Sol(x)$, $f^{-1}(w) > \frac{2}{3} \frac{1}{|Sol(x)|}$, by the proof of Theorem 3.8, $\Pr(E) > 1 - \delta$. When $E$ happens, the worst case is that $\forall w \in M_i$, $|f^{-1}(w)| = \frac{2}{3} \frac{1}{|Sol(x)|}$. Then $|M_i| \leq \frac{1}{m} \left(\frac{2}{3} \frac{1}{|Sol(x)|}\right)^{-1} = \frac{3}{2} \frac{|Sol(x)|}{m}$, and the probability that master picks $Q_i$ to dequeue is

$$\frac{L_i}{\sum_{j=1}^{m} L_j} = \frac{\sum_{w \in S_i} |f^{-1}(w)|}{\sum_{w \in S} |f^{-1}(w)|} < \frac{|S_i| \frac{2}{3} \frac{1}{|Sol(x)|}}{|S| \frac{2}{3} \frac{1}{|Sol(x)|}} = \frac{|S_i|}{|S|}.$$

Let $\Delta$ be the time between any two consecutive outputs of master. In our two-phase algorithm, we force $\Delta$ to be $(1 + \alpha) \frac{3}{2} \frac{t}{m}$, then after $T_k = k\Delta$ time, at most $k$ solutions are dequeued and output by master. Let the random variable $X_i$ be the number of times that master dequeues $Q_i$ within $T_k$, $Y$ be a random variable following the hypergeometric distribution with parameters $|Sol(x)|$, $\frac{3}{2} \frac{|Sol(x)|}{m}$ and $k$, then $\Pr(X_i \geq \frac{T_k}{t} \mid E) < \Pr(Y \geq \frac{T_k}{t})$, since $\Pr(i \text{ is picked by master}) < \frac{|S_i|}{|S|}$. Moreover, let $Z$ be a random variable following the binomial distribution with parameters $k$ and $\frac{3}{2} \frac{1}{m}$, then $\Pr(Y \geq \frac{T_k}{t}) < \Pr(Z \geq \frac{T_k}{t})$, because hypergeometric distribution is more compact than binomial distribution with the same mean [17, 19]. Since $\mathbb{E}[Z] = \frac{3k}{2m}$, by Chernoff bound [10], the probability of a queue being emptied is at most

$$\Pr\left(X_i \geq \frac{T_k}{t} \,\middle|\, E\right) < \Pr\left(Z - \mathbb{E}[Z] \geq (1 + \alpha)\frac{3}{2}\frac{k}{m} - \frac{3k}{2m}\right)$$

$$= \Pr\left(Z - \mathbb{E}[Z] \geq \alpha \mathbb{E}[Z]\right) \leq e^{-\frac{k\alpha^2}{2m}}.$$

Let $F$ be the event that no queue is emptied at any moment after $Q$ solutions are output by master. Then,

$$\Pr(\neg F | E)$$

$$= \Pr\left(\bigvee_{k=Q+1}^{|Sol(x)|} \bigvee_{i=1}^{m}\left(X_i \geq \frac{T_k}{t}\right)\middle| E\right) \leq \sum_{k=Q+1}^{|Sol(x)|} \sum_{i=1}^{m} \Pr\left(X_i \geq \frac{T_k}{t}\middle| E\right)$$

$$< m\sum_{k=Q}^{|Sol(x)|} e^{-\frac{k\alpha^2}{2m}} < m \frac{e^{-\frac{Q\alpha^2}{2m}}}{1 - e^{-\frac{\alpha^2}{2m}}} < \left(\frac{2m^2}{\alpha^2} + m\right)e^{-\frac{Q\alpha^2}{2m}}.$$

The inequality is true since $\forall x > 0, e^{-x} < \frac{1}{x+1}$. Then, $\forall \delta^* \in (0, 1)$, we have $\Pr(\neg F | E) < \delta^*$ when the parameter $Q$ is set to

$$Q \geq \frac{2m}{\alpha^2}\log\frac{2m^2 + m\alpha^2}{\alpha^2\delta^*} = \Omega\left(\frac{m}{\alpha^2}\log\frac{m}{\alpha\delta^*}\right).$$

Therefore, if the master starts enumeration after $\Theta(\frac{m}{\alpha^2}\log\frac{m}{\alpha\delta^*})$ solutions are collected in each queue (i.e., no queue is empty before $Q$ solutions are output completely), until phase-II terminates, all the $m$ queues will never be emptied with probability at least

$$\Pr(F) \geq \Pr(F|E) \cdot \Pr(E) = (1 - \delta^*)(1 - \delta) > 1 - \delta - \delta^*.$$

That is, the time between any two solutions consecutively output by master is $(1 + \alpha)\frac{3}{2}\frac{t}{m} = (1 + \alpha)\frac{3}{2}\frac{s}{m}(1 + o(1))$ with probability at least $1 - \delta - \delta^*$.

□

## B EXPERIMENTS

In Section 3, we present three algorithms, i.e., the polynomial delay random-order enumeration algorithm (PDREnum), polynomial delay Las Vegas random-order enumeration algorithm (PDLVREnum), and fully polynomial delay Atlantic City random-order enumeration algorithm (FPACREnum). In Section 4, we parallelize these algorithms on clusters.

As PDLVREnum is the core of this work, we focus on implementing and evaluating PDLVREnum and its parallelized versions for a self-reducible NP-enumeration problem admitting FPTAS, i.e., the KNAPSACK enumeration problem.

Given a knapsack with a capacity limit and a set of items each having a size. The goal of the KNAPSACK enumeration problem is to enumerate all possible packings of items such that the total size of the packed items does not exceed the knapsack's capacity limit. Mathematically, the KNAPSACK enumeration problem can be expressed as $(X_{ks}, Sol_{ks})$, where

$$X_{ks} = \{(C, s_1, \ldots, s_n) \mid n \in \mathbb{N}^+, C, s_1, \ldots, s_n \in \mathbb{N}\}$$

$$Sol_{ks}((C, s_1, \ldots, s_n)) = \{S \mid S \subseteq \mathbb{N}[1, n], \sum_{i \in S} s_i \leq C\}.$$

It is easy to observe that the KNAPSACK enumeration problem is a self-reducible NP-problem under a proper encoding scheme. Gopalan et al. [16] presented an FPTAS for the KNAPSACK enumeration problem . Then Corollary 1 follows immediately from Theorems 3.4.

COROLLARY 1. *The* KNAPSACK *enumeration problem admits an* PDLVREnum.

### B.1 Experimental Setup

Our experiments evaluate the efficiency of the algorithms, by measuring the total time and the average delay of enumerating all (or a part of) solutions.

The dataset is adopted from [6] by only keeping 20 (out of 24) items in each instance, because the time efficiency of one compared algorithm (i.e., P-SWOR(KS)) is hard to be measured on a larger dataset. We compare the following algorithms:

- PDLVREnum(KS): it is the algorithm that we design in Section 3.2.

- SWOR(KS): it is the sampling-without-replacement algorithm by Capelli and Strozecki [7] that is mentioned in the Introduction Section. It achieves random-order enumeration in each emission by simply generating a solution uniformly and outputting it only if it has not been enumerated earlier.
- P-PDLVREnum(KS): it is the parallelized version of our PDLVREnum(KS), which is a special case of the algorithms proposed in Section 4.
- P-SWOR(KS): it is a parallelized version of SWOR(KS), with the classic M/S paradigm. Each slave samples a solution uniformly and sends it to master. Master saves all solutions enumerated by slaves. To eliminate duplications, master maintains a set to store solutions that have been enumerated. A solution is output by master only if it has not been in its set.

**Implementation.** All the evaluated algorithms in our experiments are implemented with Python 3.8.8. The parallelized version of algorithms (i.e., P-LVREnum(KS) and P-SWOR(KS)) are performed on a collection of heterogeneous PCs to simulate a real-world scenario. More specifically, we set up a local cluster of six slaves, with the configurations listed in Table 1.

The other stand-alone algorithms run on a PC with Windows 10 system, the 11th generation Core i7 processor, and 32GB of RAM.

As shown in Section 3, the approximate counting algorithms are implemented and used as black boxes (i.e., oracles). For any possible prefix $w'$, we estimate $|Sol(\Psi(x, w'))|$ with brute-force-counting and simulate errors by adding random noise. The communication in the parallelized algorithms P-LVREnum(KS) and P-SWOR(KS) is implemented with the low-level networking interface (socket) in the standard library of Python.

**Dataset.** For the KNAPSACK enumeration problem, we adopt the data from [6] by only keeping 20 (out of 24) items in each instance, because the time efficiency of *P-SWOR(KS)* is hard to be measured on a larger dataset.

| Slave | CPU | RAM | Operating System |
|-------|-----|-----|------------------|
| 1 | Intel Core i7-11700 | 32GB | Windows 10 |
| 2 | Intel Core i5-8279U | 8GB | macOS Catalina |
| 3 | Intel Core i7-9750H | 32GB | Windows 11 |
| 4 | AMD Ryzen7 4800U | 16GB | Windows 11 |
| 5 | Intel Core i7-12700 | 16GB | Windows 11 |
| 6 | Intel Core i7-12750H | 32GB | Windows 10 |

**Table 1: Configurations of Slaves**

## B.2 Experiment Results

This section analyzes the efficiency of LVREnum(KS) and SWOR(KS) for the KNAPSACK enumeration problem ($X_{k_S}, Sol_{k_S}$) in the case of running on a single machine and the efficiency of the parallelized algorithms (i.e., P-PDLVREnum(KS) and P-SWOR(KS)) in the case of cluster computing.

**Single Machine.** Given an instance, we run LVREnum(KS) and SWOR(KS) on a single machine, and measure the running time and the average delay (computed by every 1000 solutions) for enumerating different amounts of solutions. As shown in the left sub-figure of Figure 1, the average delay of LVREnum(KS) remains consistently small, demonstrating the stability of our LVREnum(KS). On the other hand, the average delay of SWOR(KS) explodes quickly as the number of enumerated solutions increases. This is because the average time to generate an unenumerated solution increases exponentially for each valid emission. Eventually, SWOR(KS) becomes super inefficient after most solutions have been enumerated. Consequently, the accumulated running time of SWOR(KS) grows drastically, while that of LVREnum(KS) increases near-linearly.

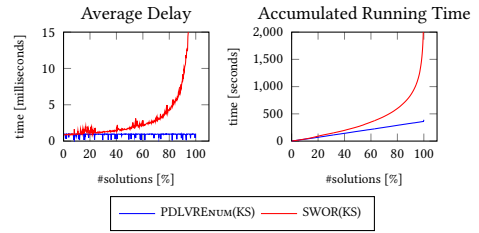

**Figure 1: The accumulated running time and the average delay (of 1000 solutions) of LVREnum(KS) and SWOR(KS) on a single machine**

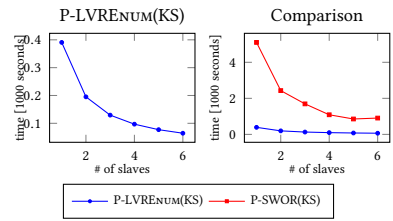

**Figure 2: The total running time of P-LVREnum(KS) on a cluster (left), and a comparison of the total running time of both algorithms (right).**

**Cluster Computing.** P-PDLVREnum(KS) and P-SWOR(KS) are the parallelized algorithms aiming at improving the efficiency of the entire enumeration process. As our parallelization is deployed on the master/slave paradigm, the improvement of efficiency is positively associated with the number of slaves. We test with $m \in \{1, 2, 3, 4, 5, 6\}$ slaves and record the total running time.

Observing from Figure 2, extra slaves do help to decrease the total running time of both algorithms. The time efficiencies of P-SWOR(KS) and P-LVREnum(KS) are not on the same level. No matter how many slaves are used, the total running time of P-SWOR(KS) is at least ten times more than that of P-LVREnum(KS).